\documentclass[aps,prl,a4paper,twocolumn]{revtex4}
\usepackage{graphics}
\usepackage{amssymb}
\usepackage{amsmath}
\usepackage{hyperref}
\usepackage{xcolor}
\usepackage{physics}
\usepackage{graphicx}
\usepackage{fancyhdr}
\usepackage{bm}
\usepackage{braket}
\usepackage{braket}
\usepackage{mathrsfs}

\usepackage{times}
\usepackage{courier}
\usepackage{bm}
\usepackage{subfig}
\usepackage{mdframed}
\usepackage{dsfont}
\usepackage{tikz}
\usetikzlibrary{shapes,arrows}
\usepackage{bbm}
\usepackage{verbatim}
\usepackage{indentfirst}
\hypersetup{
    colorlinks=true,
    linkcolor=blue,
    urlcolor=cyan,
    citecolor=cyan,}

\newcommand{\cM}{\mathcal{M}}

\newcommand{\cI}{\mathcal{I}}

\newcommand{\ModS}{\mathcal{S}}

\def\CI{{\cal I}}

\def\CS{{\cal S}}

\def\CV{{\cal V}}

\usetikzlibrary{decorations.markings}

\begin{document}

%%%%%%%%%%%%%%%%%%%%%%%%%%%%%%%%%%%%%%%%%%%%%%%%%%%%%%%%%
%%
%%               Title
%%
%%%%%%%%%%%%%%%%%%%%%%%%%%%%%%%%%%%%%%%%%%%%%%%%%%%%%%%%%

\title{Defect Relative Entropy} 

\author{Mostafa Ghasemi}
\email{ghasemi.mg@ipm.ir}
\affiliation{School of physics, Institute for Research in Fundamental Sciences (IPM) \vskip 5pt\\
	P.O.Box 19395-5531, Tehran, Iran }
	
\date{\today}

\begin{abstract}
	
Distinguishability is central to quantum information theory, but a quantitative measure for distinguishing topological defects--realizations of generalized symmetries in quantum field theory (QFT)--has been lacking. We introduce the notion of the \textit{defect relative entropy} to fill this gap for topological defects in two-dimensional conformal field theories (CFTs). For defects on a circle, we derive a universal formula that reduces defect relative entropy to a Kullback--Leibler divergence determined entirely by the modular $\mathcal{S}$-matrix and defect coefficients. Thus, the algebraic data governing modular transformations also determines defect distinguishability.
A striking consequence is that certain distinct defects can have vanishing relative entropy when restricted to one side, implying that an observer confined to that side cannot distinguish them. This gives rise to information-theoretic equivalence classes of defects, which we term \textit{defect relative sectors}. We further introduce the sandwiched defect R\'enyi relative entropy and defect fidelity, derive general formulas for these quantities. Explicit calculations in the Ising model, tricritical Ising model, and $\widehat{su}(2)_k$ WZW models illustrate our results.

%We further introduce the sandwiched defect R\'enyi relative entropy and defect fidelity, derive general formulas for these quantities, and verify our results through explicit calculations in the Ising, tricritical Ising, and $\widehat{su}(2)_k$ WZW models.

\end{abstract}

\maketitle

%%%%%%%%%%%%%%%%%%%%%%%%%%%%%%%%%%%%%%%%%%%%%%%%%%%%%%%%%
%%
%%               Contents
%%
%%%%%%%%%%%%%%%%%%%%%%%%%%%%%%%%%%%%%%%%%%%%%%%%%%%%%%%%%
\noindent\textit{Introduction}\textemdash 
Interfaces and defects play a central role in the study of quantum field theories, providing a unified framework for extended operators, generalized symmetries, and dualities~\cite{Fuchs:2002cm,Bachas:2001vj,Frohlich:2006ch,Frohlich:2009gb}. They can be regarded as natural generalizations of boundary conditions, interpolating between possibly distinct theories. Among these, \emph{conformal defects} are particularly powerful probes, as they preserve conformal symmetry and allow for exact analytical control \cite{Oshikawa:1996ww,Oshikawa:1996dj,Bachas:2007td,Quella:2006de,Frohlich:2004ef}. Their properties are encoded in the fusion of defect operators and the associated representation theory of the chiral algebra~\cite{Bachas:2007td,Frohlich:2006ch}. Of particular interest among conformal interfaces are \emph{renormalization group (RG) interfaces}, which separate two CFTs related by an RG flow. Such interfaces encode universal information about the flow and provide a non-perturbative realization of RG domain walls \cite{Gaiotto:2012np}.

A distinguished subclass is formed by \emph{topological defects}, whose correlation functions depend only on topological data and are invariant under smooth deformations of the defect line \cite{Fuchs:2007tx,Brunner:2013ota,Petkova:2000ip,Petkova:2001ag}. In rational conformal field theories (RCFTs), topological defects are classified by representations of the chiral algebra and correspond to simple objects in the associated modular tensor category~\cite{Fuchs:2002cm}. These defects implement exact symmetries, dualities, and projections onto superselection sectors.

Quantum entanglement \cite{Horodecki:2009zz} is a defining feature of quantum systems, distinguishing them from their classical counterparts by capturing intrinsically non-local correlations. A fundamental measure of entanglement is the \emph{entanglement entropy} (EE). For bipartite pure states, the EE is defined as the von Neumann entropy of the reduced density matrix $\rho_A$ associated with a spatial region $A$,
$
	S_E = -\tr (\rho_A \log \rho_A),
$
where $\rho_A$ is obtained by tracing out the degrees of freedom in the complementary region $A^{c}$.

Conformal interfaces between two CFTs can be characterized by an \emph{effective central charge}, which governs the amount of entanglement transmitted across the interface. This effective central charge does not admit a simple expression in terms of the individual central charges of the two theories; instead, it depends sensitively on the transmissive and reflective properties of the interface \cite{Sakai:2008tt,Brehm:2015lja}.

	Entanglement entropy is a powerful diagnostic of quantum correlations, but it comes with its own set of challenges. In relativistic theories, it exhibits ultraviolet divergences arising from short-distance modes near the entangling surface. Moreover, in gauge theories its definition can be ambiguous due to the non-factorization of the Hilbert space \cite{Casini:2013rba}. These difficulties motivate the use of alternative information-theoretic quantities, notably relative entropy. Although relative entropy is not a direct measure of entanglement, it is closely related to entanglement entropy, mutual information, and conditional entropy \cite{Vedral:2002zz}. Importantly, relative entropy is ultraviolet finite, universal, and free of gauge ambiguities \cite{Casini:2013rba}.

Relative entropy quantifies the distinguishability between quantum states and plays a central role in quantum information theory \cite{Vedral:2002zz} and quantum statistical mechanics \cite{Wehrl:1978zz}. Given two reduced density matrices $\rho$ and $\sigma$, the relative entropy is defined as
\begin{equation}
	D(\rho\|\sigma)
	= \mathrm{tr}(\rho \log \rho)
	- \mathrm{tr}(\rho \log \sigma).
	\label{eq:relative_entropy}
\end{equation}
It satisfies $D(\rho\|\sigma)\geq 0$, with equality if and only if $\rho=\sigma$. This definition captures
how distinguishable the two states are, with larger values indicating greater dissimilarity.

Relative entropy enjoys several key properties, including positivity and monotonicity under completely positive trace-preserving maps. These features make it a powerful tool in various areas of physics such as quantum field theory \cite{Faulkner:2016mzt, Balakrishnan:2017bjg, Casini:2019qst}, RG flow \cite{Casini:2016fgb,Casini:2022bsu,Casini:2016udt}, conformal field theory \cite{Lashkari:2014yva,Lashkari:2015dia, Ruggiero:2016khg}, boundary conformal field theory \cite{Ghasemi:2024wcq}, holography \cite{Blanco:2013joa,Wong:2013gua, Jafferis:2015del}, quantum gravity \cite{Casini:2008cr,Wall:2011hj,Longo:2018zib}, and  random states \cite{Kudler-Flam:2021rpr, Ghasemi:2024yzw}.

A one-parameter generalization of relative entropy is
known as the sandwiched R\'enyi relative entropy \cite{Tomamichel:2013bde,Wilde:2013bdg},
\begin{equation}
	D_n(\rho\|\sigma)
	= \frac{1}{n-1}
	\log
	\mathrm{tr}
	\left[
	\left(
	\sigma^{\frac{1-n}{2n}}
	\rho
	\sigma^{\frac{1-n}{2n}}
	\right)^n
	\right].
	\label{eq:renyi_relative_entropy}
\end{equation}
For $n=2$ this reduces to the collision relative entropy, while $n=\tfrac{1}{2}$ is related to fidelity.  Quantum fidelity is a useful tool for characterizing quantum phase transitions~\cite{Shi-Gu:2008zq}. The limit $n\to1$ reproduces Eq.~\eqref{eq:relative_entropy}. In practice, relative entropy can be computed via a replica trick \cite{Lashkari:2015dia},
\begin{equation}
	D(\rho\|\sigma)
	=
	-\partial_n
	\log
	\left(
	\frac{\mathrm{tr}(\rho\,\sigma^{n-1})}{\mathrm{tr}(\rho^n)}
	\right)
	\Big|_{n\to1}.
	\label{eq:replica}
\end{equation}
The concept of relative entropy can be applied to arbitrary states within spatial subregions. This involves partitioning the Hilbert space and calculating the relative entropy for density matrices in excited states $\rho$ and $\sigma$ when they are reduced to the subsystem. We apply this framework to topological defects in conformal field theories and examine how their properties depend on topological data.  Topological defect operators  $\mathcal{I}$ admit an expansion in Ishibashi-like defect projectors  $\|{\bf i}\|\equiv P^{\bf i}$ \cite{Petkova:2000ip,Petkova:2001ag}, which provide a natural basis for analyzing their algebraic and entropic structures. 
The entanglement entropy associated with an interface or defect is known as the interface (or defect) entanglement entropy (DEE) \cite{Sakai:2008tt}. This quantity has been studied in various contexts, including free boson theories \cite{Sakai:2008tt}, Ising model \cite{Brehm:2015lja, Northe:2025zmv}, conformal field theories \cite{Brehm:2015plf, Gutperle:2015kmw, Wen:2017smb, Nishioka:2021cxe, Northe:2025qcv},  symmetric product orbifolds \cite{Gutperle:2024rvo}, and holographic systems \cite{Gutperle:2015hcv, Karch:2023evr, Karch:2024udk}.

In a similar vein, we introduce the \emph{defect relative entropy} (DRE),  which is defined as the relative entropy associated with topological defect operators,  providing a measure of distinguishability in the space of defects. By utilizing the replica trick, we compute the DRE for topological defects $\mathcal{I}$ in rational conformal field theories. Our results yield a universal form for DRE that is determined by the typical data of the CFT.

For diagonal rational conformal field theories, topological interfaces are in one-to-one correspondence with primary operators labeled by $a$. In this case, the DRE between two defects $\mathcal{I}_a$ and $\mathcal{I}_{a'}$ can be expressed in terms of the modular $\mathcal{S}$-matrix as: %$D_{\CI_a,\CI_{a'}}=$
\begin{align}
	\label{IE_diag_RCFTo}
	D(\mathcal{I}_a \| \mathcal{I}_{a'})
	= \sum_{j}
	\left| \mathcal{S}_{aj} \right|^2
	\log \left|
	\frac{\mathcal{S}_{aj}}{\mathcal{S}_{a'j}}
	\right|^2 .
\end{align}
The corresponding defect fidelity is given by
\begin{equation}
	F(\mathcal{I}_a \| \mathcal{I}_{a'})
	= \sum_{j}
	\left|
	\mathcal{S}_{aj}
	\right|
	\left|
	\mathcal{S}_{a'j}
	\right| .
\end{equation}
As explicit examples, we evaluate the defect relative entropy for topological defects in the Ising model, the tricritical Ising model, and the $\widehat{su}(2)_k$ Wess--Zumino--Witten model.

In the aforementioned models, the relative entropy between the reduced density matrices of certain topological defects vanishes, even though these states are not identical. Operationally, an observer confined to one side of the entangling surface cannot distinguish topological defects related by a simple current. This leads us to define the defect relative sector as the set of topological defects with vanishing defect relative entropy. We then identify the relative entanglement sectors for these models. Our findings suggest that the defect relative sector probes a distinguished subgroup of the global symmetry. Operationally, the vanishing of defect relative entropy implies that an observer with access to only one side of the defect cannot distinguish between $\mathcal{I}_K$ and $\mathcal{I}_{K'}$ when they are related by a global symmetry, such as the $\mathbb{Z}_2$ center symmetry of $\widehat{su}(2)_k$ WZW models.

	These results establish defect relative entropy as a new quantitative probe of topological defect structure, complementary to fusion rules and modular data. Our companion paper applies this framework to symmetric orbifold CFTs, where KL divergence decomposes into permutation-group and modular contributions~\cite{Ghasemi:2026sij}.

\textit{ Defect relative entropy}\textemdash
In this work, we focus on rational conformal field theories (RCFTs), whose Hilbert space decomposes as
\begin{equation}
	\mathcal{H}
	=
	\bigoplus_{i,\bar j}
	\mathcal{M}_{i\bar j}
	\,
	\mathcal{V}_i \otimes \bar{\mathcal{V}}_{\bar j},
\end{equation}
where $\mathcal{V}_i$ and $\bar{\mathcal{V}}_{\bar j}$ are irreducible representations of the chiral and anti-chiral Virasoro algebras, and $\mathcal{M}_{i\bar j}$ are finite multiplicities.

An operator corresponding to a general topological interface  between two RCFTs with multiplicity matrices $\cM_{i\bar j}^{(1)}$ and $\cM_{i\bar j}^{(2)}$  can be then  written as $\CI_K = \sum_{\bf i}\,d_{K\bf i}\,P^{\bf i}$~\cite{Petkova:2000ip,Brehm:2015plf}.
Here, $K$ labels types of topological interface and ${\bf i} \equiv (i, \bar j; \alpha,\beta)$ is the index for the Ishibashi-type  projector $P^{\bf i}$ intertwining between two  representations:
\begin{align}
P^{\bf i} : \; \left(\CV_i\otimes \bar \CV_{\bar j}\right)^{(\alpha)} \;\longrightarrow\; \left(\CV_i\otimes \bar \CV_{\bar j}\right)^{(\beta)} \ .
\end{align}
Here, $(i,\bar{j})$ labels the transmitted pair of representations. The indices $\alpha = 1, \dots, \mathcal{M}_{i\bar{j}}^{(1)}$ and $\beta = 1, \dots, \mathcal{M}_{i\bar{j}}^{(2)}$ denote the multiplicity labels of this pair on the two sides of the interface.
These interfaces fuse by composition: $d_{KK'\mathbf{i}} = d_{K\mathbf{i}} \, d_{K'\mathbf{i}}$. Modular invariance imposes the ``Cardy'' condition~\cite{Petkova:2000ip}
\begin{equation}\label{CARD-COND}
	\sum_{\mathbf{i}} S_{ij} S_{\bar i\bar j} \; \operatorname{Tr}\!\left(d_{K^*\mathbf{i}} \, d_{K\mathbf{i}}\right) = \mathcal{N}_{\mathbf{j}K}^{\;K} \in \mathbb{N}_0,
\end{equation}
which restricts admissible coefficients and defines elementary interfaces as indecomposable ones. Here $\operatorname{Tr}$ denotes the trace over multiplicity indices $\alpha,\beta$, treating $d_{K(i,\bar j;\alpha,\beta)}$ as a matrix, with the convention $d_{K^*(i,\bar j;\alpha,\beta)} \equiv d_{K(i,\bar j;\beta,\alpha)}^*$. For time evolution parallel to the interfaces, the coefficient $\mathcal{N}_{\mathbf{j}K'}^{\;K}$ counts the multiplicity of the representation pair $(j,\bar{j})$ in a configuration where topological interfaces intersect a spatial slice.
%The coefficient $\mathcal{N}_{\mathbf{j}K'}^{\;K}$ counts the multiplicity of the representation pair $(j,\bar{j})$ in a configuration where topological interfaces intersect a spatial slice, i.e., for time evolution parallel to the interfaces.

 For two isomorphic CFTs, where $\cM_{i\bar j}^{(1)} = \cM_{i\bar j}^{(2)}$, the projector $P^{\bf i}$ has a realization as
\begin{align}
	P^{(i, \bar j; \alpha, \beta)} \equiv \sum_{ {\bf n}, \bar {\bf n}}\, \left(|i,  {\bf n}\rangle\otimes |\bar j, \bar  {\bf n}\rangle\right)^{(\alpha)}\, \left(\langle i,  {\bf n}|\otimes \langle \bar j, \bar  {\bf n}|\right)^{(\beta)} \ ,
\end{align}
where $|i, {\bf n}\rangle\otimes |\bar j, \bar {\bf n}\rangle$
denotes an orthogonal basis for the representation $\CV_i\otimes \bar\CV_{\bar j}$.
In this case, we have the following commutation relations with the Virasoro generators $	[L_n, \CI_K] = [\bar L_n, \CI_K] = 0 $. 
%Defects intertwining between the same theory are called interfaces.

To compute the defect relative entropy $D(\mathcal{I}_K\|\mathcal{I}_{K'})$ between two arbitrary topological interfaces $\mathcal{I}_K$ and $\mathcal{I}_{K'}$, we employ the replica trick in the thermodynamic limit ($\ell/\epsilon \gg 1$). Following the replica approach for entanglement entropy across a topological interface~\cite{Sakai:2008tt,Brehm:2015plf,Gutperle:2015kmw}, the $n$th replica partition function reduces to a torus partition function with $2n$ defect insertions of the topological interface operators $\mathcal{I}_K$ and $\mathcal{I}_{K'}$ . This approach yields \footnote{see Supplemental Material \ref{SUPLEM:DT-DRE}.},
\begin{equation}
	G_n(\mathcal{I}_K\|\mathcal{I}_{K'}) = \frac{Z_n(K,K')\,(Z_1(K))^{n-1}}{Z_n(K)\,(Z_1(K'))^{n-1}}.
\end{equation}
%The DRE is then $D(\cI_K\|\cI_{K'}) = -\, \partial_n \log G_{n}(\cI_K\|\cI_{K'})\Big|_{n \to 1}$. Evaluating this gives our main result:
The DRE is then $D(\mathcal{I}_K\|\mathcal{I}_{K'}) = -\,\partial_n \log G_n(\mathcal{I}_K\|\mathcal{I}_{K'})\big|_{n\to 1}$, and evaluating this gives our main result:
\begin{align}\label{KUL-RE}
	D(\cI_K\|\cI_{K'}) = 
	\sum_{(i, \bar j)}\,\Tr\left[ p^K_{\bf i}\,\log  \frac{p^K_{\bf i}}{p^{K'}_{\bf i}}\right] \ .
\end{align}
 The $p^K_{\bf i}$ is a probability distribution characterized by the modular $\CS$-matrix and defect coefficients as
\begin{align}
	p^K_{\bf i} &\equiv \frac{\mathcal{S}_{0i}\,\left(\mathcal{S}_{0\bar{j}}\right)^\ast}{\sum_{(i,\bar j)}\,\mathcal{S}_{0i}\,\left(\mathcal{S}_{0\bar{j}}\right)^\ast\,\Tr \left[ d_{K {\bf i}}\,d_{K^\ast {\bf i}}\right]}\,d_{K {\bf i}}\,d_{K^\ast {\bf i}}  \ .
\end{align}
 Eq.~\eqref{KUL-RE} can be interpreted as the Kullback--Leibler (KL) divergence~\cite{KullbackLeibler} between the probability distributions associated with $\mathcal{I}_K$ and $\mathcal{I}_{K'}$ on the $\mathrm{CFT}_1$ side. This quantity is positive and provides a key measure of distinguishability in the space of topological defects.

Similarly, the sandwiched defect R\'enyi relative entropy takes the form
\begin{equation}\label{SNRE-1}
	D_n(\mathcal{I}_K\|\mathcal{I}_{K'}) = \frac{1}{n-1}\log\sum_{(i,\bar j)}\operatorname{Tr}\!\left[ (p^K_{\mathbf{i}})^{n}\, (p^{K'}_{\mathbf{i}})^{1-n} \right].
\end{equation}
For $n=1/2$, this relates to the \textit{defect fidelity}
\begin{equation}
	F(\mathcal{I}_K\|\mathcal{I}_{K'}) = \sum_{(i,\bar j)}\operatorname{Tr}\!\left[ \sqrt{p^K_{\mathbf{i}}}\,\sqrt{p^{K'}_{\mathbf{i}}}\right],
\end{equation}
which generalizes the pure-state overlap $|\langle\phi|\psi\rangle|$ to topological defects and reduces to the classical fidelity between probability distributions.

For diagonal theories $( \cM_{i\bar j} = \delta_{i\bar j})$, the CFTs on both sides are the same theory and topological interfaces are in one-to-one correspondence with primary operators labeled by $a$: %$\CI_a = \sum_i\,\frac{\mathcal{S}_{ai}}{\mathcal{S}_{0i}}\,P^{i} $,
\begin{align}\label{Topological_Interface}
	\CI_a = \sum_i\,\frac{\mathcal{S}_{ai}}{\mathcal{S}_{0i}}\,P^{i}, \quad P^{i} \equiv \sum_{{\bf n}, \bar {\bf n}}\, |i, {\bf n}\rangle\otimes |\bar i, \bar {\bf n}\rangle\, \langle i, {\bf n}|\otimes \langle \bar i, \bar {\bf n}|  \ ,
\end{align}
where $P^{i} $ denotes the projector acting on the representation $\CV_i\otimes \bar\CV_{\bar i}$.
Then the probability distribution simplifies to $	p^a_i = \left|\mathcal{S}_{ai}\right|^2$, and DRE and DF takes the form:
\begin{align}\label{IE_diag_RCFT}
D(\mathcal{I}_a \| \mathcal{I}_{a'}) &=\sum_{j} \left| \mathcal{S}_{aj}\right|^2 \,\log \left|\frac{\mathcal{S}_{aj}}{\mathcal{S}_{a'j}} \right|^2.%-  \sum_{j} \S_{lj}^2 \log \left( \frac{\S_{kj}^2}{\S_{1j}} \right).
\end{align}
and  
\begin{equation}\label{DF_diag_RCFT}
	F(\CI_a\|\CI_{a'}) = \sum\limits_{j} \left|\mathcal{S}_{aj}\right|\, \left|\mathcal{S}_{a'j}\right|.
\end{equation} 
Equation~(\ref{IE_diag_RCFT}) shows that $D(\mathcal{I}_a\|\mathcal{I}_{a'})$ diverges whenever $|\mathcal{S}_{a'j}|^2 = 0$ for some $j$ with $|\mathcal{S}_{aj}|^2 \neq 0$; otherwise, it is positive \cite{Ghasemi:2024wcq}. 
The defect relative entropy can also vanish. This naturally leads to the definition of a \textit{defect relative sector}, comprising those topological defects for which $D(\mathcal{I}_a\|\mathcal{I}_{a'}) = 0$.

The vanishing of defect relative entropy admits a natural operational interpretation. Since this quantity is defined using data accessible from one side of the defect, defects in the same relative sector are indistinguishable to an observer restricted to that side. This is consistent with the monotonicity of relative entropy under restriction of the observable algebra.

 Remarkably, Eqs.~(\ref{IE_diag_RCFT}) and~(\ref{DF_diag_RCFT}) coincide exactly with the left-right relative entropy and left-right fidelity of Ref.~\cite{Ghasemi:2024wcq}. This follows from the one-to-one correspondence between topological defects and boundary states in diagonal RCFTs, which share identical expansion coefficients. Consequently, the \textit{relative entanglement sector} of~\cite{Ghasemi:2024wcq} is isomorphic to the defect relative sector defined here.

\textit{Examples}\textemdash
%\emph{Examples.---}
In this section, we will illustrate specific examples to demonstrate how to calculate the defect relative entropy. Equation \eqref{IE_diag_RCFT} demonstrates that having the modular $\mathcal{S}$ matrix enables the direct computation of the DRE and DF for the defects $\mathcal{I}_a$. We will explicitly compute the DRE and DF for three models: the Ising model and the  $\widehat{su}(2)_{k}$  WZW model. The tricritical Ising model is computed in Appendix.,

Ising model: The critical Ising CFT ($c = 1/2$) is equivalent to a $\mathbb{Z}_2$ orbifold of a free massless Majorana fermion CFT via 2D bosonization.
It contains  three primary operators : the identity $\mathbb{I}$, the thermal operator $\epsilon$, and the spin field $\sigma$
with conformal weights of $
0 , \ 1/2 , \ 1/16
$, respectively. 
The modular $\mathcal{S}$ matrix for the Ising model can be found in \cite{yellowbook}.
The three elementary topological defects of the Ising model are $\CI_{\mathrm{id}}$, $\CI_{\epsilon}$, and $\CI_{\sigma}$.
The defect corresponding to the vacuum is the identity defect  $\CI_{\mathrm{id}}$. The defect $\CI_{\epsilon}$ is a symmetry defect implementing the $\mathbb{Z}_2$ symmetry of the Ising model. The presence of these two defects does not result in a shift of the entanglement entropy. The third defect $\CI_{\sigma}$ implements Kramers-Wannier duality. It satisfies the fusion rules
$
\CI_{\sigma} \CI_{\sigma} = \CI_{\mathrm{id}} + \CI_{\epsilon}. 
$
When the vacuum defect $\CI_{\mathrm{id}}$ is taken as the reference state, the defect relative entropy for the Ising model is determined by the following expression:
\renewcommand{\arraystretch}{1.5}
\begin{center}
	\begin{tabular}{  c l }
		\hline
		\ \ \ \ Topological defects\ \ \ \                                        & \ \ \ \ \ \ \  \ \ DRE \ \ \ \ \ \ \ \ \ \ \   DF \ \ \ \\ \hline\hline
		$\CI_{\epsilon}$ \quad $\CI_{\mathrm{id}}$ &\ \ \ \ \ \ \ \ \ \ \ \ \ $ 0\ \ \ \  $ \quad  \ \ \ \ \ \ \ $ 1\ \ \ \  $ \\
		$\CI_{\sigma}$ \quad  $\CI_{\mathrm{id}}$           & \ \ \ \ \ \ \ \ \ \ \ \ \ $ \log 2 \ $    \quad  \ \ \ $ \frac{\sqrt{2}}{2}\ \ \ \  $
		\\ \hline
	\end{tabular}
\end{center} \null 
For the critical Ising model, the defect relative sector—defined as the set of topological defects with vanishing relative entropy—are $E_1 = \{\CI_{\mathrm{id}}, \CI_{\epsilon}\}$ and $E_2 = \{\CI_{\sigma}\}$.

\def\suck{\widehat{su}(2)_{k}}
The $\widehat{su}(2)_{k}$ WZW model:
The $\suck$ WZW model has a  central charge of $c=3k/(k+2)$ and at level $k$ has irreducible representations are labeled by spins \( j \), where \( j = 0, \frac{1}{2}, 1, \ldots, \tfrac{k}{2}\). In this model, there is a $\mathbb{Z}_2$ center symmetry that commutes with the left and right current algebras and acts on the representation labels as $j \mapsto \tfrac{k}{2} - j$, corresponding to a nontrivial automorphism of the fusion ring.
Using the theory's modular $\mathcal{S}$ matrix~\cite{Fendley:2006gr} and  equation (\ref{IE_diag_RCFT}), we obtain the DRE for the diagonal mass matrix $\mathcal{M}=\mathbb{I}$ as:
\begin{equation}\label{DRE-WZW}
	D_{j,j'} = \sum_{l=0}^{k/2} p_j(l) \log \left( \frac{p_j(l)}{p_{j'}(l)} \right),
\end{equation}
where $p_j(l) = \frac{2}{k+2} \sin^2 \theta_{jl}$ and $\theta_{jl} \equiv \frac{\pi (2j+1)(2l+1)}{k+2}$. 
The defect relative  sectors of $\widehat{su}(2)_{k}$ WZW  models consist of states with zero DRE. This condition holds if and only if:  $j = j' \quad \text{or} \quad j = \tfrac{k}{2} - j'$, reflecting $\mathbb{Z}_2$ symmetry equivalences among topological defects.
	
	For DRS sectors defined by \([\CI_j] = \{\CI_j, \CI_{\tfrac{k}{2} - j}\}\), even \(k\) admits a unique fixed point at \(j = k/4\), yielding a non-anomalous singlet, while all other sectors form \(\mathbb{Z}_2\) doublets. For odd \(k\), all sectors are doublets with no fixed point. 	This level-dependent pattern parallels the presence or absence of a \(\mathbb{Z}_2\) 't Hooft anomaly in boundary states at odd and even levels, respectively~\cite{Numasawa:2017crf}. This pattern is also observed in the relative entanglement sector~\cite{Ghasemi:2024wcq}.

For example, let's set $k=2$. Topological defects in the  $\widehat{su}(2)_{2}$ WZW model correspond to primary fields with spins $
j = 0, \frac{1}{2}, 1$,
These topological defects are organized under a \(\mathbb{Z}_2\) symmetry as follows:
A symmetric pair of topological defects
$
\{\CI_{\mathrm{id}}, \CI_{1}\},
$
and a single \(\mathbb{Z}_2\) invariant topological defect
$
\CI_{ \frac{1}{2}}.
$
\begin{center}
	\begin{tabular}{  c l }
		\hline
		\ \ \ \ Topological defects\ \ \ \                                        & \ \ \ \ \ \ \ \ \ \ DRE \ \  \ \ \ \ \ \   DF \ \ \   \\ \hline\hline
		$\CI_1$ \quad $\CI_{\mathrm{id}}$ &\ \ \ \ \ \ \ \ \ \ \ \ \ $ 0\ \ \ \  $ \quad  \ \ \ \ \ \ \ $ 1\ \ \ \  $ \\
		$\CI_{ \frac{1}{2}}$ \quad  $\CI_{\mathrm{id}}$           & \ \ \ \ \ \ \ \ \ \ \ \ \ $ \log 2 \ $    \quad  \ \ \ $ \frac{\sqrt{2}}{2}\ \ \ \  $    
		\\ \hline
	\end{tabular}
\end{center} \null 
The defect relative sectors are \( \{\CI_{\mathrm{id}}, \cI_1\} \) and \( \{\CI_{ \frac{1}{2}}\} \). This structure realizes the Tambara-Yamagami (TY) fusion category \( \mathrm{TY}_{\mathbb{Z}_2}^- \)~\cite{Choi:2023xjw}. Similar computations of DRE can be carried out for higher-rank WZW models or coset constructions.

Regarding the vanishing of the defect relative entropy between two reduced density matrices, this phenomenon occurs exclusively when the corresponding topological defects are related by a global symmetry (e.g., the $\mathbb{Z}_2$ center symmetry in $\widehat{su}(2)_k$ WZW models). Operationally, an observer confined to one side of the entangling surface cannot distinguish topological defects that are related by such a symmetry.

\textit{Discussion}\textemdash
%\emph{Discussion.---}
In this work, we have introduced defect relative entropy as a measure of distinguishability for topological defects in CFTs on a circle. The defect relative entropy reduces to a Kullback--Leibler divergence built from the modular $\mathcal{S}$-matrix and defect coefficients, and it vanishes for some class of topological distinct defects that form information-theoretic equivalence classes—defect relative sectors. This establishes a direct link between defect distinguishability and modular data, and opens the door to an information-theoretic classification of generalized symmetries.

Here, we have calculated the defect relative entropy for topological defects. It would be interesting to calculate it for conformal defects as well.

%It would be interesting to investigate defect relative entropy along defect renormalization group (RG) flow. As a measure of distinguishability, it may provide a new information-theoretic probe of defect RG flows and offer insights into possible monotonicity properties. 

%It would also be worthwhile to study the evolution of defect relative sectors under RG flow and to explore whether their structure is related to infrared dualities or emergent equivalences between defects.

%An interesting direction for future work would be to identify the gravitational dual of defect relative entropy in the AdS/ICFT context.

It would be interesting to investigate defect relative entropy along defect renormalization group (RG) flows. As a measure of distinguishability, it may provide a new information-theoretic probe of such flows and offer insights into possible monotonicity properties. It would also be worthwhile to study the evolution of defect relative sectors under RG flow and to explore whether their structure is related to infrared dualities or emergent equivalences between defects.

 An additional direction for future work would be to identify the gravitational dual of defect relative entropy within the AdS/ICFT framework.

%An interesting direction for future work is to study defect relative entropy in the context of defect RG flows~\cite{...}. Beyond serving as a quantitative measure of distinguishability along the flow, it may provide a useful probe of the evolution of defect relative sectors. Understanding whether such sectors are preserved or reorganized under RG evolution could shed light on infrared dualities and emergent equivalences among defects.

%\appendix

%\section{Positivity of the LRRE, Eq. 19}
%Here, we prove the positivity of LRRE, Eq. 19, as a measure of distinguishability.\\

	\begin{acknowledgments}
	\emph{Acknowledgements.---}
	It is my pleasure to thank  Sepideh Mohammadi for  reading our manuscript and providing insightful comments.% We are sincerely grateful to ....... for his valuable comments and insightful suggestions.
	
\end{acknowledgments}

\bibliographystyle{utphys}

%\endgroup

%%%%%%%%%%%%%%%%%%%%%%%%%%%%%%%%%%%%%%%%%%%%%%%%%%%%%%%%%%%
%                     END MATTER                           %
%%%%%%%%%%%%%%%%%%%%%%%%%%%%%%%%%%%%%%%%%%%%%%%%%%%%%%%%%%%

\section*{Supplemental Material: Detailed Derivations and More Example}
\label{SUB-MAT}

\subsection*{SM1.  Detailed Derivations of Defect Relative Entropy}
\label{SUPLEM:DT-DRE}
The computation of defect relative entropy relies on the replica trick, which was employed in Refs.~\cite{Sakai:2008tt,Brehm:2015plf,Gutperle:2015kmw} to compute entanglement entropy across a topological interface. Within this approach, the entanglement entropy is obtained from the partition function $Z_n$ on an $n$-sheeted Riemann surface $\mathcal{R}_n$, which is equivalent to a torus partition function with $2n$ insertions of the defect operator $\mathcal{I}_K$~\cite{Sakai:2008tt,Brehm:2015plf,Gutperle:2015kmw}.

Similarly, the defect relative entropy is obtained from the $n$th replica partition function. Using Eq.~\eqref{eq:replica}, the $n$th R'enyi relative entropy is determined by
$
\operatorname{Tr}\!\big(\rho_K \, \rho_{K'}^{\,n-1}\big).
$
Unlike the entanglement entropy case, the insertion of different density matrices $\rho_K$ and $\rho_{K'}$ explicitly breaks the replica $\mathbb{Z}_n$ symmetry. Equivalently, this trace can be mapped to a torus partition function with $2n$ insertions of the topological interface operators $\mathcal{I}_K$ and $\mathcal{I}_{K'}$. This leads to the following expression:
\begin{align}
	G_{n}(\mathcal{I}_K\|\mathcal{I}_{K'})
	&= \frac{Z_{n}(K,K')\,(Z_{1}(K))^{n-1}}{Z_{n}(K)\,(Z_{1}(K'))^{n-1}} .
\end{align}
For the general topological interface  $\CI_K = \sum_{\bf i}\,d_{K\bf i}\,P^{\bf i}$, the corresponding partition functions defined as,
\begin{widetext}
	\begin{align}
		\begin{aligned}\label{DEF-PAR1}
			Z_{n}(K,K')&=  \tr\left[ \left( \CI_K\, e^{-t H}\, \CI_K^\dagger\,e^{-t H}\right)\left( \CI_{K'}\, e^{-t H}\, \CI_{K'}^\dagger\,e^{-t H}\right)^{(n-1)}\right]\\
			&= \tr\left[ \left(\CI_K\,\CI_K^\dagger\right)\left(\CI_{K'}\,\CI_{K'}^\dagger\right)^{(n-1)}\,e^{-2t n H}\right] \\
			&= \sum_{(i,\bar j)}\, \Tr\left[\left(d_{K {\bf i}}\,d_{K^\ast {\bf i}}\right)\,\left(d_{K' {\bf i}}\,d_{K'^\ast {\bf i}}\right)^{(n-1)}\right]\, 
			\chi_i \left(e^{-2nt}\right)\, \chi_{\bar j} \left(e^{-2nt}\right) \ ,
		\end{aligned}
	\end{align}
\end{widetext}
and
\begin{align}
	\begin{aligned}\label{DEF-PAR2}
		Z_{n}(K) 
		&= \tr\left[ \left( \CI_K\, e^{-t H}\, \CI_K^\dagger\,e^{-t H}\right)^n\right]\\
		&= \tr\left[ \left(\CI_K\,\CI_K^\dagger\right)^n\,e^{-2t n H}\right] \\
		&= \sum_{(i,\bar j)}\, \Tr\left[\left(d_{K {\bf i}}\,d_{K^\ast {\bf i}}\right)^{n}\right]\, \chi_i \left(e^{-2nt}\right)\, \chi_{\bar j} \left(e^{-2nt}\right) \ ,
	\end{aligned}
\end{align}
where $H= L_0 + \bar L_0 - \frac{c}{12}$ is the Hamiltonian on a cylinder. In the above equations, we used the commutation relation  $	[L_n, \CI_K] = [\bar L_n, \CI_K] = 0 $ in the second equality.
$\chi_i$ denotes the character in the $\CV_i$ representation and
the parameter $t$ is related to the UV and IR cutoffs $\epsilon$, $L$ as $t = \frac{2\pi^2}{\log ( L/\epsilon)}$.
$\text{Tr}$ denotes the trace over multiplicity indices $\alpha, \beta$ by regarding $d_{K(i,\bar j;\alpha, \beta)}$ as a matrix with the notation $d_{K^\ast (i,\bar j;\alpha, \beta)} \equiv d_{K(i,\bar j;\beta, \alpha)}^\ast$. 
Using the relations (\ref{DEF-PAR1}) and (\ref{DEF-PAR2}), and modular transformation property  of character $ \chi_i(e^{-2nt}) = \sum_j \ModS_{ij} \, \chi_j\!\left(e^{-\frac{4\pi^2}{2nt}}\right)$, we obtain the following expression
in the thermodynamic limit $\ell/\epsilon \gg 1$,
\begin{align}\label{KUL-RE1}
	D(\cI_K\|\cI_{K'}) & = -\, \partial_n \log G_{n}(\cI_K\|\cI_{K'})\Big|_{n \to 1} \nonumber\\&
	=
	\sum_{(i, \bar j)}\,\Tr\left[ p^K_{\bf i}\,\log  \frac{p^K_{\bf i}}{p^{K'}_{\bf i}}\right]. \,
\end{align}
where $p^K_{\bf i}$ is a probability distribution characterized by the modular $\CS$-matrix and defect coefficients as
\begin{align}
	p^K_{\bf i} &\equiv \frac{\mathcal{S}_{0i}\,\left(\mathcal{S}_{0\bar{j}}\right)^\ast}{\sum_{(i,\bar j)}\,\mathcal{S}_{0i}\,\left(\mathcal{S}_{0\bar{j}}\right)^\ast\,\Tr \left[ d_{K {\bf i}}\,d_{K^\ast {\bf i}}\right]}\,d_{K {\bf i}}\,d_{K^\ast {\bf i}} \ , 
	\nonumber\\&
	p^\text{Id}_{\bf i} \equiv \mathcal{S}_{0i}\,\left(\mathcal{S}_{0\bar{j}}\right)^\ast\,\delta_{\alpha\beta} \ .
\end{align}
For each $(i,\bar j)$, $p_{\mathbf{i}}^{A}$ is a positive-semidefinite Hermitian matrix \footnote{In unitary theories, $\mathcal{S}_{i0} > 0$.}, so its eigenvalues are real and non-negative and satisfy $\sum_{(i,\bar j)} \operatorname{Tr} p_{\mathbf{i}}^{A} = 1$, thereby forming a probability distribution. In a quantization where time is orthogonal to the interface, $\operatorname{Tr} p_{\mathbf{i}}^{A}$ gives the probability of finding system CFT$_1$ in the Ishibashi-type state labeled by $(i,\bar j)$ after tracing out CFT$_2$; thus $\{p_{\mathbf{i}}^{A}\}$ defines a reduced density matrix \cite{Brehm:2015plf}.

The Eq.(\ref{KUL-RE1}) is the general formula for the defect relative entropy and is free from any cut-off dependent divergence for any choice of $\CI_K$, as we naturally expect from relative entropy.

%......................................................

%In CFT$_1$ we define a probability distribution from the topological interface coefficients:
%$p_{(\mathbf{i},\alpha\alpha')}^{A} = (d_{A^*\mathbf{i}}d_{A\mathbf{i}} S_{i0}S_{\bar\imath}0)/\mathcal{N}_{\mathbf{0}A}^{\;A}$,
%where $\mathbf{i}=(i,\bar\imath;\alpha,\alpha')$ and $\alpha,\alpha'$ run over multiplicities $M_{i\bar\imath}^1$.
%For each $(i,\bar\imath)$, $p_{\mathbf{i}}^{A}$ is a positive-semidefinite Hermitian matrix, and
%$\sum_{(i,\bar\imath)}\operatorname{Tr} p_{\mathbf{i}}^{A}=1$,
%so its eigenvalues form a probability distribution. In a quantization with time orthogonal to the interface,
%$\operatorname{Tr} p_{\mathbf{i}}^{A}$ gives the probability of finding CFT$_1$ in the Ishibashi-type state labeled by $(i,\bar\imath)$
%after tracing out CFT$_2$; thus $\{p_{\mathbf{i}}^{A}\}$ defines a reduced density matrix.
%For the identity defect, $p_{\mathbf{i}}^{\text{id}} = S_{i0}S_{\bar\imath}0\,\delta_{\alpha\alpha'}$.

\subsection*{SM2. Worked Example: Tricritical Ising model ($c=7/10$)}

The tricritical Ising model is the Virasoro minimal model $\mathcal{M}(5,4)$ with $c=7/10$. This theory contains six primary operators
with conformal dimensions
$
0  , \  {1}/{10}   , \ {3 / 5}   , \ {3/ 2}   , \ {3 / 80}   
$ and $ {7 / 16} $. And,  associated to each of these primary operators,
there is a topological defect.
The results for the DRE, for various topological defects, are listed in the following table. (Here, $s_1=\frac{1}{\sqrt{5}}\sin\frac{2\pi}{5}$ and $s_2=\frac{1}{\sqrt{5}}\sin\frac{4\pi}{5}$. These appear as entries in the modular $\mathcal{S}$ matrix \cite{yellowbook,Castro:2011zq}).
\begin{widetext}
	\renewcommand{\arraystretch}{1.5} 
	\begin{center}
		\begin{tabular}{  c l } 
			\hline
			Topological defects\ \ & \ \ \ \ \ \ \ \ \ \ \ \ \ \ \ \ \ \ \ \ \ \ \ \ \ \ \ \ \ \ \ \ \ \   \ \ \ \ DRE \ \   \\ 
			\hline
			\hline
			$\CI_\frac{3}{2}$ \quad $\CI_{\mathrm{id}}$ &\ \ \ \ \ \ \ \ \ \ \ \ \ \ \ \ \ \ \ \ \ \ \ \ \ \ \ \ \ \ \ \ \ \ \ \ \ \ \ \ $ 0  $\\
			$\CI_\frac{1}{10}$ \quad $\CI_{\mathrm{id}}$ &\ \ \ \ \ \ \ \ \ \ \ \ \ \ \ \ \ \ \ \ \ \ \ \ \ \ \ \ \ \ \ \ \ \ \ \ \ \ \ \ $ 8({\left(\mathit{s}_1^2- \mathit{s}_2^2\right) \left( \log \mathit{s}_1\right)+\left(\mathit{s}_2^2- \mathit{s}_1^2\right) \left(\log\mathit{s}_2  \right)})  $\\
			$\CI_\frac{3}{5}$ \quad $\CI_{\mathrm{id}}$ &\ \ \ \ \ \ \ \ \ \ \ \ \ \ \ \ \ \ \ \ \ \ \ \ \ \ \ \ \ \ \ \ \ \ \ \ \ \ \ \ $ 8({\left(\mathit{s}_1^2- \mathit{s}_2^2\right) \left( \log \mathit{s}_1\right)+\left(\mathit{s}_2^2- \mathit{s}_1^2\right) \left(\log\mathit{s}_2  \right)})  $\\
			$\CI_\frac{7}{16}$ \quad $\CI_{\mathrm{id}}$ & \ \ \ \ \ \ \ \ \ \ \ \ \ \ \ \ \ \ \ \ \ \ \ \ \ \ \ \ \ \ \ \ \ \ \ \ \ \ \ \ $8\mathit{s}_1^2\log\left(\frac{\mathit{s}_1}{\mathit{s}_2}\right) +4\left(\mathit{s}_2^2- \mathit{s}_1^2\right)\log2 $
			\\
			$\CI_\frac{3}{80}$ \quad $\CI_{\mathrm{id}}$ & \ \ \ \ \ \ \ \ \ \ \ \ \ \ \ \ \ \ \ \ \ \ \ \ \ \ \ \ \ \ \ \ \ \ \ \ \ \ \ \ $ 4\left(\mathit{s}_1^2+ \mathit{s}_2^2\right)\log2 $
			\\
			\hline
		\end{tabular}
	\end{center}
\end{widetext}
In this model, the non-trivial defect relative sectors are $\{\CI_{\mathrm{id}},\CI_\frac{3}{2}\}$ and $\{\CI_\frac{3}{5},\CI_\frac{1}{10}\}$.

\end{document}